\documentstyle[aps,multicol,psfig]{revtex}  

\newcommand\beq{\begin{equation}} 
\newcommand\eeq{\end{equation}} 
\newcommand\bea{\begin{eqnarray}} 
\newcommand\eea{\end{eqnarray}} 
\newcommand\bB{{\bf B}} 
\newcommand\bE{{\bf E}} 
\newcommand\be{{\bf e}} 
\newcommand\bK{{\bf k}} 
\newcommand\bx{{\bf x}} 
\newcommand\br{{\bf r}} 
 
\begin{document}    
\title{  
Correlations of electromagnetic fields in chaotic cavities 
}    
\author{B. Eckhardt, U. D\"orr, U. Kuhl and H.-J. St\"ockmann}    
\address{Fachbereich Physik, Philipps Universit\"at   
	Marburg, D-35032 Marburg, Germany}   
    

\maketitle
 
\begin{abstract}    
We consider the fluctuations of electromagnetic fields in 
chaotic microwave cavities. We calculate the transversal 
and longitudinal correlation function based on a random 
wave assumption and compare the predictions with measurements 
on two- and three-dimensional microwave cavities. 
\end{abstract}    
	 
\begin{multicols}{2}

Classical ergodicity suggests that wave functions in chaotic systems 
may be described by superpositions of waves with wave vectors of 
constant length but random directions \cite{Voros,Ber77a}.  
Their fluctuations 
are distinctly different from the more familiar optical speckle patterns 
where also the wave numbers fluctuate \cite{Ocon87}. 
The distributions of amplitudes turns out to be Gaussian and the  
spatial autocorrelation function is given by Bessel functions of order  
$\frac{d}{2}-1$, where $d$ is the billiard dimension.  
Overwhelming evidence for this has been accumulated especially in  
numerical studies of billiards \cite{McD79,McD88,Shapiro}, 
experiments on microwave billiards \cite{Kud95,Pri95b} 
and surfaces of constant negative  
curvature \cite{Aur93}. Higher order moments and correlations  
\cite{Sre96a,Sre96b} as well 
as contributions from prominent classical structures 
(`scars') have also been studied \cite{Hell_scar}, so that by now 
one has a fairly good understanding of the fluctuation properties 
of scalar wave functions in chaotic systems. 
 
Recent experiments on the elastodynamics of vibrating blocks  
\cite{ell96}  
and 3-d microwave resonators  
\cite{doe98,Alt97}  
deal with situations where the wave fields have more than one  
component which are typically mixed by the boundary conditions 
\cite{baldup,fra96,Ott,Weaver}. 
New effects such as ray splitting \cite{Ott} 
or chaotic features 
in systems with integrable ray dynamics \cite{Weaver} are found.  
 
We are here interested in the consequences of these additional 
degrees of freedom for the fluctuations of the wave functions. 
In particular, we will show that the fact that in the  
absence of space charges 
electromagnetic fields are divergence free implies 
differences between the longitudinal and transversal correlation 
functions and deviations from the behaviour expected  
for scalar fields~\cite{Ber77a}. We present results for the 
field components, the intensities and the frequency shift  
and compare with experiments on microwave billiards. 
 
Our starting point is the electromagnetic analog of the 
the semiclassical ansatz that the scalar field is a superposition 
of plane waves with constant amplitudes and fixed wave length but  
randomly oriented wave vector \cite{Ber77a}.  
For the electromagnetic case we have 
in addition to allow for different orientations of the polarization. 
We thus assume that the field at a point  
${\bf x}$ in position space is due to  
a superposition of many plane waves with  
uniformly distributed orientations of polarization and wave vector, e.g., 
\beq 
\bE(\bx) \approx \sum_\nu \bE_\nu  
e^{i\bK_\nu\cdot\bx}\,, 
\label{E_ansatz} 
\eeq 
and similarly for the $\bB$-field.  
The complex amplitudes $\bE_\nu$ and  
$\bB_\nu$ of all waves are transversal, 
$\bE_\nu\cdot\bK_\nu=0$ and $\bB_\nu\cdot\bK_\nu=0$, 
and satisfy  
$\bE_\nu\cdot\bB_\nu=0$  
so that $\bK_\nu$, $\bE_\nu$ and $\bB_\nu$ form an orthogonal 
dreibein. 
The absolute values  
$|\bE_\nu|$ and $|\bB_\nu|$ are all the same, that is to say, we  
assume that there are no losses during reflections at the 
walls. From this assumption there 
follows immediately that all three components 
are Gaussian distributed with the same  
distribution~\cite{doe98}
 
The spatially averaged correlation functions then 
are for the electric field 
\beq 
C_{E,ij}(\br)= \langle\bE_i(\bx+\br/2) \, \bE_j(\bx-\br/2) \rangle 
/\langle {\bf E}_i^2\rangle \,, 
\eeq 
and similar for the magnetic field and the cross correlation between 
$\bE$ and $\bB$. The normalization is by the mean square amplitude of 
a single component of the fields, $\langle E_i^2\rangle$, so that 
$C_{E,ii}(0)=1$. 
In a tensor notation, this can be combined to a tensor 
of correlation functions, 
\beq 
C_{E}(\br)= \langle\bE(\bx+\br/2) \otimes \bE(\bx-\br/2) \rangle 
/\langle {\bf E}_i^2\rangle \,. 
\eeq 
Substituting (\ref{E_ansatz}) and performing the spatial average  
then results in (the normalization will be restored in the end) 
\beq 
C_{E}(\br) \propto 
\sum_\nu \bE_\nu \otimes \bE_\nu^* 
e^{i\bK_\nu\cdot \br} \,. 
\eeq 
To proceed further, let $\br$ point in the $z$-direction and introduce 
spherical coordinates for the wave vector, 
\beq 
\bK_\nu = k \left( 
\matrix{\cos\phi_\nu\, \sin\theta_\nu \cr 
	\sin\phi_\nu\, \sin\theta_\nu \cr 
	\cos\theta_\nu}\right)\,. 
\eeq 
The electromagnetic field contributions lie in a 
plane perpendicular to this wave vector, spanned by 
the two vectors 
\beq 
\be_1^{(\nu)} = \left( 
\matrix{-\sin\phi_\nu \cr 
	\cos\phi_\nu \cr 
	0}\right)\, 
\,,\quad 
\be_2^{(\nu)} =  \left( 
\matrix{-\cos\phi_\nu\, \cos\theta_\nu \cr 
	-\sin\phi_\nu\, \cos\theta_\nu \cr 
	\sin\theta_\nu}\right) \,. 
\eeq 
If $\psi_\nu$ denotes the angle of polarization, the  
field components are 
\bea 
\bE_\nu &=& \cos\psi_\nu\, \be_1^{(\nu)} + \sin\psi_\nu\, \be_2^{(\nu)} \\ 
\bB_\nu &=& - \sin\psi_\nu\, \be_1^{(\nu)} + \cos\psi_\nu\, \be_2^{(\nu)}\,. 
\eea 
In the limit of a large number of contributing components, the  
sum over the different contributions can be replaced by a continuous 
average over all directions  
(angles $\theta$ and $\phi$) for the wave vector and all polarizations 
(angle $\psi$), 
\beq 
\frac{1}{N}\sum_\nu \cdots \rightarrow 
\frac{1}{2\pi} \int_0^{2\pi} d\psi\, 
\frac{1}{2\pi} \int_0^{2\pi} d\phi\, 
\frac{1}{2} \int_0^{\pi} \sin\theta\,d\theta\, \cdots 
\eeq 
After averaging over the polarizations and the  
azimuthal angle, the correlation functions become 
\bea 
C_{E}(\br) &=& \frac{3}{4} \left(\matrix{1 & 0 & 0 \cr 
		      0 & 1 & 0 \cr 
		      0 & 0 & 0 }\right) 
		      \langle e^{ikr\cos\theta}\rangle_\theta \cr 
	&+& \left\langle \left(\matrix{\frac{3}{4}\cos^2\theta & 0 & 0 \cr 
		      0 & \frac{3}{4}\cos^2\theta& 0 \cr 
		      0 & 0 & \frac{3}{2}\sin^2\theta }\right) 
		      e^{ikr\cos\theta}\right\rangle_\theta\,. 
\eea 
The final average over $\theta$ can be  
expressed in terms of spherical Bessel functions,  
but it is more convenient to use the trigonometric representation 
directly, 
\beq 
C_{E}(\br) = \left(\matrix{f_\perp(kr) & 0 & 0 \cr 
		      0 & f_\perp(kr) & 0 \cr 
		      0 & 0 & f_\parallel(kr) }\right) 
\eeq 
with the transversal correlation function 
\beq 
f_\perp(\xi) = \frac{3}{2} \left({\sin \xi \over\xi } -  
{\sin \xi - \xi \cos\xi  \over \xi^3}\right) 
\label{f_t} 
\eeq 
and the longitudinal correlation function 
\beq 
f_\parallel(\xi) =  3 { \sin\xi  - \xi \cos \xi \over \xi^3}\,. 
\label{f_l} 
\eeq 
The asymptotic behaviour of these functions is that for small 
$r$ they approach the same value, $C_{E, ii}\rightarrow 1$, by 
normalization.  
For large $r$ they oscillate on a scale set by the wavenumber 
and decay like $1/r$ for the transversal 
and like $1/r^2$ for the longitudinal correlations. 
For the trace of the correlation function, 
\bea 
\mbox{tr\,} C_{E} &=& \langle \bE(\bx+\br/2) \cdot\bE(\bx-\br/2) 
\rangle/\langle {\bf E}_i^2\rangle \\ 
&=& 3 {\sin kr \over kr} 
\label{f_tr} 
\eea 
the correlation between polarizations and wave vector is eliminated 
and one  
recovers Berry's result for random waves in three dimensions \cite{Ber77a}, 
except for a factor due to the normalization 
(see Fig.~1). 
 
\begin{figure}  
\narrowtext
 \centerline{ 
 \psfig{figure=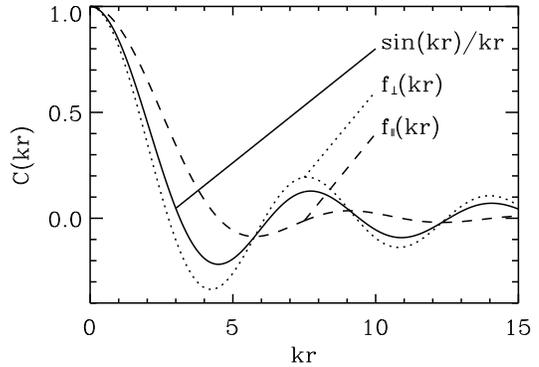,angle=0,width=8cm}  } 
 \caption[]{
 Correlation functions for longitudinal and transversal fluctuations 
 (dashed and dotted line, respectively) for the electromagnetic fields. 
 The full line shows the correlation function for 
 a 3-d scalar wave function for comparision.} 
\end{figure} 

The correlations for the magnetic field $\bB$ have the same 
functional dependence, 
\beq 
C_{B}(\br) = \left(\matrix{f_\perp(kr) & 0 & 0 \cr 
		      0 & f_\perp(kr) & 0 \cr 
		      0 & 0 & f_\parallel(kr) }\right)\,. 
\eeq 
There are no correlations between $\bE$ and $\bB$. 
 
For the experiments also the correlations of intensities, 
\beq 
C_{EE}(\br) = \langle |\bE(\bx+\br/2)|^2\,  
|\bE(\bx-\br/2)|^2 \rangle 
/\langle |{\bf E}|^4\rangle\, 
\eeq 
and similarly for the magnetic field are relevant. For ease of 
comparison with numerical data, we normalize this function by the 
second moment of the intensity so that $C\rightarrow 1$ as  
$\br \rightarrow 0$. For large $\br$ the correlations between intensities 
decay and the correlation function approaches  
$ \langle |{\bf E}|^2\rangle^2 / \langle |{\bf E}|^4\rangle$. 
For Gaussian random field components, this ratio is $3/5$. Therefore, 
the correlation function with the above normalization becomes 
\beq 
C_{EE}(\br) = \frac{4}{15}\left[f_\perp(kr)\right]^2 +  
\frac{2}{15}\left[f_\parallel(kr)\right]^2+\frac{9}{15} 
\label{C_EE} 
\eeq 
and similar for the magnetic field.  
For the intensity correlation function of a 3-d scalar field  
the correlation function becomes 
\beq 
C_{ss}(\br) = \frac{2}{3}\left({\sin kr\over kr}\right)^2 +\frac{1}{3}\,, 
\label{C_ss} 
\eeq 
where the asymptotic value of $1/3$ reflects the ratio of  
the square of the second moment to the fourth moment as 
expected for a Gaussian distribution \cite{Sre96a,Sre96b}. 
 
 \begin{figure}  
\narrowtext
 \centerline{ 
 \psfig{figure=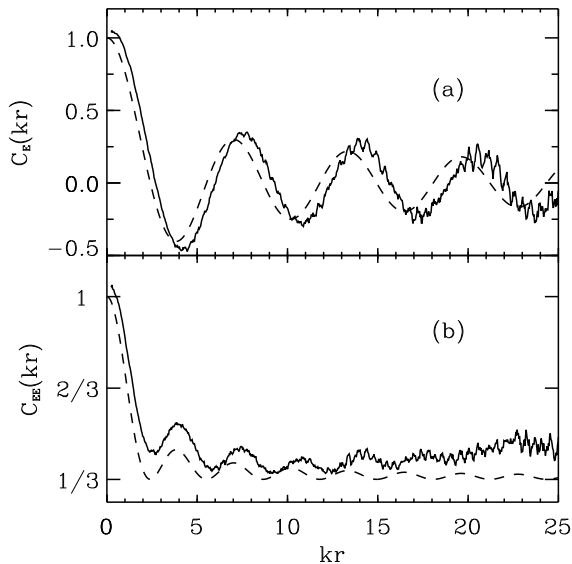,width=8cm}  } 
 \caption[]{\label{intensities} 
 Comparison between experimental results for spatial autocorrelation  
 functions in a quarter stadium microwave billiards and theory.  
 The experimental curves were obtained by  
 superimposing the results for the 20 lowest lying eigenresonances.  
 (a) Spatial autocorrelation function of the electric field amplitudes 
 (full line) and theoretical predictions from Eq.~(\ref{2-d-corr}).  
 (b)  
 Spatial autocorrelation function for the squares of the electric field  
 amplitudes using the same data set as in (a). The dashed line corresponds  
 to the prediction from Eq.~(\ref{2-d-intens}). 
 } 
\end{figure} 

To test these predictions we measured the field distributions in two- and  
three-dimensional microwave billiards. We start with the discussion of the  
results in a resonator of the shape of a quarter stadium billiard. We  
measured the microwave transition amplitudes between two antennas, one  
kept fixed, the other moved around to probe the spatial distribution 
of the wave functions. At an eigenfrequency such  
measurements yield directly the electric field strength $\bE(x)$  
as a function of the position. Details of the experiment are  
described elsewhere~\cite{Ste95}. For microwave frequencies below  
$\nu_{max}=c/2d$, where $d$ is the height of the resonator,  
only TM modes are excited and the electric field has a single 
component $E_z$. For this component, Berry's arguments give 
a spatial autocorrelation function 
\beq 
C_E(\br)= \langle E_z(\bx+\br/2) \cdot E_z(\bx-\br/2)\rangle 
\sim J_0(kr) 
\label{2-d-corr} 
\eeq 
Fig.~2(a) shows the experimental autocorrelation function, normalized to  
$C_E(0)=1$. It was obtained by superimposing the results from the 20 lowest  
eigenfrequencies of the quarter stadium. Apart from a discrepency of about  
10 percent in the wavelength of the oscillations the experiment 
reproduces the prediction by Eq.~(\ref{2-d-corr}) perfectly. 
 
The autocorrelation function of the field intensities becomes
in the 2-d case 
\bea 
C_{EE}(\br) &=& \langle |E_z(\bx+\br/2)|^2\, |E_z(\bx-\br/2)|^2 \rangle 
\nonumber\\ &\sim&\frac{2}{3}\left[J_0(kr)\right]^2+\frac{1}{3} 
\,. 
\label{2-d-intens} 
\eea 
This correlation function,
which has also been studied by Sridhar et al. \cite{Kud95},
 is shown in  
Fig.~2(b) for the same data set that entered Fig.~2(a).   
As in Fig.~2(a) we note a small difference in the wavelength of the 
oscillations. This has not been observed in the experiments of 
Sridhar et al.~\cite{Kud95} and in the stadium wave functions  
in Ref.~\cite{McD88}, but it has appeared 
for wave functions of an octogon billiard on a surface with constant  
negative curvature \cite{Aur93}, where it has been
attributed to anisotropies in the wave functions.
In the present case the discrepancy may be caused 
by higher order corrections to the semiclassical 
predictions since the  
wave functions are not very far into the semiclassical regime: 
Typical wave lengths for the wave functions that entered the 
analysis of Fig. 2 are about 0.2--0.5 stadium widths. 
Assuming corrections to be of order $1/k^2$ with a prefactor of  
order 1, as suggested by the analysis of \cite{Gas93}, the deviations 
can be estimated to be about 10\%, as observed. 
 
 \begin{figure}
 \narrowtext  
 \centerline{ 
 \psfig{figure=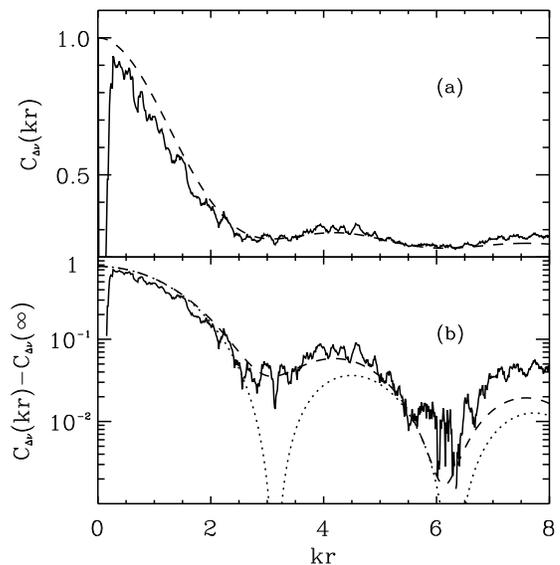,width=8cm}  } 
 \caption[]{\label{experiment} 
 Spatial autocorrelation function $C_{\Delta\nu}(kr)$ for the frequency  
 shifts obtained by the perturbing bead method in a three-dimensional Sinai  
 billiard. The dashed line corresponds to the theoretical prediction from  
 Eq.~(\ref{C_nu}),  
 taking into account the vector properties of the electromagnetic  
 fields. (a) Correlation function on a linear scale. (b) The  
 correlation function $C_{\Delta\nu}(kr)-C_{\Delta\nu}(\infty)$  
 on a logarithmic plot. The dotted  
 line corresponds to Eq.~(\ref{C_ss}) and would have been expected for  
 three-dimensional scalar fields.} 
\end{figure} 

Turning to the three-dimensional microwave cavities note that 
Maxwell's equations can no longer be  
reduced to a scalar wave equation, so that effects due to the  
vector properties of the electromagnetic field can become essential.  
The field distributions in a cavity of the shape of a three-dimensional  
Sinai billiard were mapped by means of the perturbing bead  
method~\cite{doe98}. The technique uses the fact that a  
spherical metallic bead in the resonator shifts the  
eigenfrequency of a resonance by an amount  
$\Delta\nu$ proportional to $-2\bE^2+\bB^2$,  
where $\bE$ and $\bB$ are the fields at the position of the bead.  
Since the electric and the magnetic field components are uncorrelated,  
the spatial autocorrelation function for the frequency shift, 
\beq 
C_{\Delta\nu}(r)=\langle \Delta\nu(\bx+\br/2)  
\cdot\Delta\nu(\bx-\br/2)\rangle 
\label{C-nu} 
\eeq 
is given by the intensity  
autocorrelation function defined in Eq.~(\ref{C_EE}), up to 
an off-set resulting  
from the fact that $C_{\Delta\nu}(r)$ does not vanish in the limit  
$r\to\infty$. This off-set is easily calculated. Using again that for  
Gaussian distributions the fourth moment amounts to three times the square  
of the second moment, we find 
$C_{\Delta\nu}(0)/C_{\Delta\nu}(\infty)=13/3$. After  
normalization we thus have 
\beq 
C_{\Delta\nu}(r)= \frac{20}{39}\left[f_\perp(kr)\right]^2 +  
\frac{10}{39}\left[f_\parallel(kr)\right]^2 + \frac{3}{13} 
\label{C_nu} 
\eeq 
The experimental results in Fig.~3(a) are in good agreement with  
the theoretical prediction from Eq.~(\ref{C_nu}) for $kr$-values below 6.  
The hole in the histogram for $kr<0.2$ reflects the minimum grid size used  
in the bead measurement. In Fig.~3(b) the data for   
$C_{\Delta\nu}(r)-C_{\Delta\nu}(\infty)$ are replotted on a  
logarithmic scale 
to emphasize the minima. Also shown is the scalar correlator (\ref{C_ss}).  
Note that 
the scalar function has zeroes but the spectral correlator does not. The 
absence of zeroes or, in the case of finite resolution, deep minima, 
in the experimental data is thus clear evidence for the influence of the  
polarizations.  
 
In summary, we have shown that longitudinal and transversal
correlation functions of electromagnetic waves in chaotic
cavities differ because the waves are transversal.
The experiments verified the effect for the intensities, and
more direct studies of the fields themselves are desirable.
We expect that also in other situations with non-scalar waves fields,
as for instance in acoustics in anisotropic media or hydrodynamic
waves, the correlations will be characterized by tensors which
depend on the character of the modes and the directions.

This work was partially supported by the  
Deutsche Forschungsgemeinschaft through 
Sonderforschungsbereich 185 'Nichtlineare Dynamik'. 


\end{multicols}
\end{document}